\documentclass[usenatbib]{mnras}
\usepackage{graphicx}
\usepackage{multicol}	
\usepackage[english]{babel}
\usepackage{epstopdf}
\usepackage[T1]{fontenc}
\usepackage{ae,aecompl}
\usepackage{hyperref}
\usepackage{amsmath}    % Advanced maths commands
\usepackage{amssymb}    % Extra maths symbols

\def\mpc{\,h^{-1}{\rm Mpc}}
\def\kpc{\,h^{-1}{\rm kpc}}

\def\msun{\,h^{-1}{\rm M}_\odot}
\def\mstar{\,h^{-2}{\rm M}_\odot}

\usepackage{color}
\newcommand{\adb}[1]{\textcolor{blue}{ #1}} % additions in blue
\makeatletter

\newcommand{\Rmnum}[1]{\expandafter\@slowromancap\romannumeral #1@}
\makeatother

\title[galaxy-halo size relation]{Galaxy-halo size relation from Sloan Digital Sky Survey Data Release 7 and the 
ELUCID simulation}

\author[Youcai Zhang]{Youcai Zhang$^{1}$\thanks{yczhang@shao.ac.cn},
Xiaohu Yang$^{2,3}$,
Hong Guo$^{1}$
\\
$^{1}${Key Laboratory for Research in Galaxies and Cosmology,
  Shanghai Astronomical Observatory; Nandan Road 80, Shanghai 200030,
  China} \\
$^{2}$Department
of Astronomy, School of Physics and Astronomy, and Shanghai Key Laboratory for Particle Physics and Cosmology, \\
~~Shanghai Jiao Tong University, Shanghai 200240, China \\
$^{3}$Tsung-Dao Lee Institute and Key Laboratory for
    Particle Physics, Astrophysics and Cosmology, Ministry of Education, \\
    ~~Shanghai Jiao Tong University, Shanghai 200240, China
}

\begin{document}
\label{firstpage}
\pagerange{\pageref{firstpage}--\pageref{lastpage}}
\maketitle

\begin{abstract}

Based on galaxies in the Sloan Digital Sky Survey Data Release 7 (SDSS DR7) and dark matter haloes in the 
dark matter only, cosmological and constrained ELUCID simulation, we investigate the relation between 
the observed radii of central galaxies with stellar mass $\gtrsim 10^{8} \mstar$ and the virial radii of 
their host dark matter haloes with virial mass $\gtrsim 10^{10.5} \msun$, and the dependence of galaxy-halo
size relation on the halo spin and concentration. Galaxies in observation are matched to dark matter (sub-)haloes in 
the ELUCID simulation using a novel neighborhood subhalo abundance matching method. For galaxy 2D half-light 
radii $R_{50}$, we find that early- and late-type galaxies have the same power-law index 0.55 with $R_{50} \propto
R_{\rm vir}^{0.55}$, although early-type galaxies have smaller 2D half-light radii than late-type galaxies
at fixed halo virial radii. When converting the 2D half-light radii $R_{50}$ to 3D half-mass radii $r_{1/2}$,
both early- and late-type galaxies display similar galaxy-halo size relations with 
$\log r_{1/2} = 0.55 \log (R_{\rm vir}/210 \kpc) + 0.39$. We find that the galaxy-halo size ratio $r_{1/2}/ R_{\rm vir}$
decreases with increasing halo mass. At fixed halo mass, there is no significant dependence of
galaxy-halo size ratio on the halo spin or concentration.

\end{abstract}


\section{Introduction}\label{sec_intro}

In the standard galaxy formation model, galaxies form in the center of the dark matter haloes \citep{White1978}.
Therefore, the properties of galaxies are expected to be influenced by the properties of their
host dark matter haloes. The physical connection between galaxies and haloes is an interesting and 
challenging problem in the study of galaxy formation and evolution.

From the observational data, such as the Sloan Digital Sky Survey \citep[SDSS;][]{Abazajian2009} and
the Galaxy And Mass Assembly survey \citep[GAMA;][]{Driver2011}, 
it's well established that the sizes of galaxies depend not only on the luminosity and 
stellar mass, but also on other intrinsic properties of galaxies, such as the galaxy concentration or 
morphology types \citep{Shen2003, Lange2015, Zhang2019}. Unlike the mass-size relation of galaxies in 
observation \citep{VanderWel2014}, the scaling relations between galaxies and dark matter haloes 
are not well understood \citep{Dutton2009, Wechsler2018, Zanisi2020, Rohr2022}, since haloes 
are more than one order of magnitude larger than galaxies.

The angular momentum conservation model \citep{Fall1980,Mo1998} has been widely used to set the galaxy sizes
in various semi-analytic models \citep{Cole2000,Croton2006,Some2008a}  and to explain the empirical constraints
of the galaxy-halo size relation by the abundance matching technique \citep{Kravtsov2013,Some2018}. Using the 
abundance matching method, \citet{Kravtsov2013} investigated the galaxy-halo size relation for hundreds of galaxies,
which include dwarf galaxies from the HST and VLT/FORS1 observation and $220$ massive galaxies from the
SDSS observation, spanning more than eight order of magnitude in stellar mass ($M_* \sim 10^{5-12} M_{\odot}$). 
They found that the galaxy-halo size relation follows an approximately linear relation 
$r_{\rm 1/2} = 0.015 R_{200}$, where $r_{\rm 1/2}$ is the galaxy 3D half-mass radius and $R_{200}$ is the radius
enclosing the over-density of $200$ times the critical density $\rho_{\rm cirt}$ of the universe.
Combining the sizes of galaxies ($M_* > 10^7 M_\odot$) from the CANDELS survey with the inferred sizes 
of dark matter haloes by the abundance matching method with four different stellar mass-halo mass relations, 
\citet{Huang2017} confirmed the linear galaxy-halo size relation found by  
\citet{Kravtsov2013} at $z=0$, and extended to the redshift range $0< z < 3$ . In addition, they found that
early- and late-type galaxies follow a roughly parallel galaxy-halo size relation offset by 
$\sim 0.2$-$0.3$ dex. Combining galaxies ($M_* \gtrsim 10^9 M_\odot$) in the GAMA survey and subhaloes from the 
Bolshoi–Planck simulation by subhalo abundance matching
method, \citet{Some2018} claimed that the ratio of galaxy 3D half-mass radius to halo virial radius  
is consistent with being roughly independent of the galaxy stellar mass, with a linear relation 
$r_{\rm 1/2} = 0.018 R_{\rm vir}$ for galaxies from the GAMA survey at $z=0.1$. It is noteworthy that in 
their Figure 3, the ratio of galaxy projected half-light radius (or 3D half-mass radius without including scatter) 
to the halo virial radius decreases significantly with the increase of the galaxy stellar mass, although 
this trend is mitigated due to the projection effect from projected to 3D size or the including scatter 
in the stellar-to-halo mass relation in their subhalo abundance matching method. Using galaxies
($M_* \geq 10^{9.75}$) in SDSS DR10 and subhaloes in Bolshoi–Planck simulation, \citet{Hearin2019} 
studied the size dependence of galaxy 
clustering and found that small galaxies cluster more strongly than large galaxies of the same stellar mass.
They claimed that the size dependence of galaxy clustering can be well reproduced based on a simple 
linear galaxy-halo size relation with $R_{50} = 0.01 R_{\rm vir,peak}$, where $R_{50}$ is the 
galaxy 2D half-light radius, and $R_{\rm vir,peak}$ is the halo virial radius when halo mass reached its
maximum over the merger history. Using galaxies ($M_* \gtrsim 10^8 M_\odot$) from the SDSS DR7, 
\citet{Rodriguez2021} combined galaxy sizes and stellar masses with their group finder algorithm
to construct the relations between galaxy sizes and halo masses, and compared with predictions from 
the IllustrisTNG hydrodynamical simulation.  They demonstrated that the results of central galaxies 
are in excellent agreement with the linear galaxy-halo size relation proposed by \citet{Kravtsov2013}, 
while for satellite galaxies, the results are more consistent with the model of \citet{Hearin2019}.

In the angular momentum conservation model, the galaxy size is assumed to be determined by the angular 
momentum of its host halo and the galaxy spin is strongly correlated with that of its host halo, resulting
in $r_{1/2} \propto \lambda R_{\rm vir}$. However, recent studies based on cosmological hydrodynamical
simulations challenge the angular momentum conservation model and indicate that there is almost no 
correlation between the spins of a galaxy and its host halo \citep{Desmond2017, Jiang2019, YangH2021, Stiskalek2022}. 
Generally, these numerical studies suggest that the baryon physics of cooling and feedback process is more 
significant in setting galaxy sizes. Using data from the EAGLE hydrodynamical simulation, \citet{Desmond2017} 
found that the sizes of galaxies ($M_* > 10^9 M_\odot$) only weakly correlate with halo spin or 
concentration at fixed stellar mass with the Spearman rank coefficients $0.17$ and $-0.19$, which are 
consistent with 0 within $3\sigma$.
Using $15$ Milky Way-mass galaxies in the hydrodynamical, zoom-in FIRE-2 simulation, \citet{Garrison2018}
claimed that galaxy sizes are poorly correlated with the halo spin, formation time or merger histories. They found
that the spin of the gas at the time the galaxy formed half of its stars is the the best predictor of galaxy 
size. Using $34$ galaxies with halo mass $M_{\rm vir} \sim 2 \times  10^{11-12} M_\odot$ 
and $\sim 100$ galaxies with $M_{\rm vir} \sim 10^{9.5-12.3} M_\odot$ from zoom-in cosmological hydrodynamical 
VELA and NIHAO simulations, \citet{Jiang2019} found 
that the galaxy spin is barely correlated with the spin of its host halo. \citet{YangH2021} also found that there is almost no 
correlation between galaxy size and the halo spin parameter for dwarfs in the EAGLE and APOSTLE simulations,
although they claimed that there is a correlation for Milky Way sized galaxies in IllustrisTNG simulation
(see their Figure 4). Using low-mass galaxies ($M_* \sim 10^{7-9} M_\odot$) in hydrodynamical 
simulation in the FIRE project, \citet{Rohr2022}
found that the scatter of the galaxy-halo size relation does not correlate with the halo spin or concentration.

These numerical studies show that galaxy sizes are weakly correlated with halo spin and concentration, 
which disagrees with the angular momentum conservation model, although \citet{Rodriguez-Gomez2022} 
claimed that there is a strong correlation between galaxy size and halo spin in the IllustrisTNG hydrodynamical 
cosmological simulation \citep[see also][]{Grand2017}.
Based on the abundance matching method, 
\citet{Zanisi2020} examined their models in setting galaxy sizes from their host haloes, including
the MMW model in \citet{Mo1998}, the empirical model in \citet{Kravtsov2013} and the 
concentration model $r_{1/2} = 0.02(c/10)^{-0.7} R_{\rm vir}$ in \citet{Jiang2019}. 
They found that the angular momentum conservation model can not provide a good 
fit to the size distributions of SDSS galaxies. They suggested that the galaxy-halo size relation can
be mediated by the {\tt galaxy} angular momentum, not the halo spin parameter. 

In general, the galaxy-halo size relation has not been well understood based on the cosmological 
hydrodynamical simulation or the abundance matching method. 
In a series of recent studies, with the help of dark matter only, cosmological and constrained 
ELUCID simulation, the mass and positions of dark matter haloes of galaxy groups in observation
can be well produced in the ELUCID simulation, since the initial condition of the ELUCID simulation is 
constrained by the density field of galaxy distribution
in observation \citep{WangHuiyuan2012, WangHuiyuan2014, WangHuiyuan2016, Zhang2021a, Zhang2021b}. Using
the ELUCID simulation, galaxies in observation can be well linked to dark matter (sub-)haloes in simulation 
according to their mass and positions by a neighborhood subhalo abundance matching method \citep{Yang2018}.
In this work we further investigate the galaxy-halo size relation for {\it observed} central galaxies linked to the corresponding
haloes in the ELUCID simulation. In particular, we aim to provide a empirical constraint of the galaxy-halo size 
relation.

This paper is organized as follows. In Section~\ref{sec_data}, we describe the observational galaxy sizes from
SDSS DR7 and the halo properties from the ELUCID simulation. Besides, we describe the neighborhood subhalo 
abundance matching method linking galaxies in observation to haloes in constrained simulation. In 
Section~\ref{sec_result}, we investigate the relation between the observed radii of central galaxies in SDSS 
DR7 and the virial radii of their host haloes in the ELUCID simulation. In addition, we study how the 
galaxy-halo size relation depends on the halo spin and concentration. Finally, we summarize our results in
Section~\ref{sec_summary}. Throughout the paper, the adopted cosmological parameters are $\Omega_{\rm m} = 0.258$,
$\Omega_{\rm b} = 0.044$, $\Omega_\Lambda = 0.742$, $h = 0.72$, $\sigma_8 = 0.796$, and $n_{\rm s} = 0.963$.

\section{data and method}\label{sec_data}

The observational data used in this paper is from the spectroscopic galaxy sample of the Sloan Digital Sky Survey 
Data Release 7 \citep[SDSS DR7;][]{Abazajian2009}, from which \citet{Blanton2005} constructed the the New York 
University Value-Added Galaxy Catalog (NYU-VAGC) with an independent set of improved calibrations. From NYU-VAGC,
we collect a total of $639,359$ galaxies with redshifts in the range of $0.01 \le z \le 0.2$, and with 
extinction-corrected apparent magnitude brighter than $17.72$. The stellar masses of galaxies are computed from
the relation between mass-to-light ratio and the color of \citet{Bell2003}. In addition, we have checked our 
results using the stellar masses of galaxies from the public catalog of \citet{Chang2015}. The relevant results
are found to be not sensitive to the different stellar mass used. 

\begin{figure}
\includegraphics[width=0.5\textwidth]{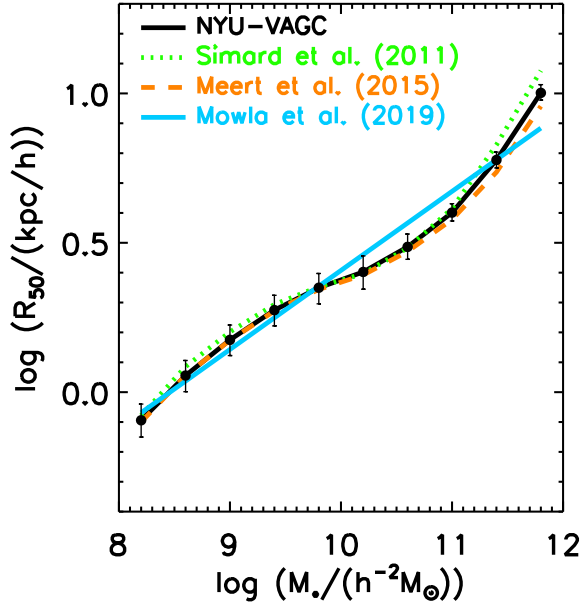}
\caption{Galaxy r-band half-light radius as a function of the stellar mass from SDSS DR7. The black solid line shows 
the median values of galaxy sizes from NYU-VAGC catalog \citep{Blanton2005}. The error bars show the $40$ per cent 
and $60$ per cent quantiles. For comparison, the green dotted and orange dashed lines show the mass-size relation 
using galaxy sizes from \citet{Simard2011} and \citet{Meert2015}, respectively. The cyan solid line shows the galaxy 
size-stellar mass relation proposed by \citet{Mowla2019}.}
\label{fig:sm_galaxy}
\end{figure}

In the NYU-VAGC catalog, the $r$-band galaxy radii $R_{50}$ and $R_{90}$ enclosing $50\%$ and $90\%$ of the Perosian flux
are extracted from the table {\tt object\_sdss\_imaging.fits}\footnote{\href{http://sdss.physics.nyu.edu/vagc-dr7/vagc2/}{\tt http://sdss.physics.nyu.edu/vagc-dr7/vagc2/}}.
For comparison, the galaxy sizes are also extracted from the public catalogs of \citet{Simard2011} and 
\citet{Meert2015}, respectively. The catalog of \citet{Simard2011} consists of $1,123,718$ galaxies fitted 
with three different galaxy profiles. We only adopt the $r$-band half-light radius in their canonical
fitting model (see {\tt Table 1} of \citet{Simard2011} for details). The catalog of \citet{Meert2015}
consists of $670,722$ galaxies fitted with four different galaxy profiles.  We adopt the $r$-band
half-light radius in the model-independent measurements fitted by the PYMORPH software pipeline
(see {\tt Table 2} of \citet{Meert2015} for details). 

From the catalogs of \citet{Simard2011} and \citet{Meert2015}, $580,472$ galaxies can be cross identified
in the NYU-VAGC catalog. Figure~\ref{fig:sm_galaxy} shows the $r$-band half-light radius as a function of 
the stellar mass for the cross matched galaxies. The black solid line shows the median values
of galaxy sizes from NYU-VAGC catalog. The blue dotted and red dashed lines show the size-mass relation 
using galaxy sizes from \citet{Simard2011} and \citet{Meert2015}, respectively. 
The cyan solid line shows the size-mass fitting relation provided by \citet{Mowla2019}, who used
$918$ massive galaxies with mass $M_* > 2\times 10^{11} M_\odot$ at $0.1 < z <3.0 $ from the COSMOS-DASH and 
HST ACS/WFC surveys, and combined them with $30,958$ galaxies with mass $M_* > 10^{9} M_\odot$ from the 
3D-HST/CANDELS sample of \citet{VanderWel2014}. As shown in 
Figure~\ref{fig:sm_galaxy}, at fixed stellar mass the galaxy sizes in NYU-VAGC catalog agree well with 
those in the catalogs of \citet{Simard2011} and \citet{Meert2015}, especially for low-mass galaxies. 
Throughout the paper, the galaxy sizes from the NYU-VAGC catalog are used unless stated otherwise.
In this paper, we mainly focus on a sample of $276,463$ central galaxies with stellar mass 
$M_* \gtrsim 10^{8} \mstar$, corresponding to haloes with the virial mass $M_{\rm vir} 
\gtrsim 10^{10.5} \msun $ in the ELUCID simulation. In this sample, the median values of galaxy stellar
mass and the corresponding halo virial mass are $10^{10.3} \mstar$ and  $10^{11.8} \msun$, respectively.

\subsection{Simulation data}

The ELUCID simulation \citep{WangHuiyuan2014, WangHuiyuan2016, Tweed2017, 
WangHY2018, Yang2018, Chen2019} is a dark matter only cosmological constrained
simulation, which contains $3072^3$ 
dark matter particles within a box that is $500 \mpc$ on a side, and has a particle mass of
$3.1 \times 10^8 \msun$. The gravitational softening length is $3.5 \kpc$. The cosmological
parameters adopted in the ELUCID simulation are $\Omega_{\rm m} = 0.258$, $\Omega_{\rm b} = 0.044$, 
$\Omega_\Lambda = 0.742$, $h = 0.72$, $\sigma_8 = 0.796$, and $n_{\rm s} = 0.963$.

Dark matter haloes are identified using the standard friends-of-friends (FOF) algorithm
\citep{Davis1985} with a linking length of $b=0.2$ times the mean particle separation. 
Using the SUBFIND algorithm \citep{Springel2001}, a given host halo is decomposed into 
a set of gravitational bound subhaloes, in which the most massive one is called central
subhalo, and the others are referred to satellite subhaloes.

For a given host halo, the virial mass $M_{\rm vir}$ is computed as the interior mass 
within a sphere of radius $R_{\rm vir}$ centered on the position of the minimum potential particle, 
where $R_{\rm vir}$ is the radius within which the over-density is $\Delta_{\rm vir}$ times 
the critical density $\rho_{\rm crit}$ of the universe. Then the viral mass and
radius are related by 
\begin{equation}\label{eqn:radius_mass}
M_{\rm vir} = \frac{4}{3} \pi  \Delta_{\rm vir} \rho_{\rm crit} R_{\rm vir}^3,
\end{equation}
where $\Delta_{\rm vir}$ is dependent on the cosmological parameters and given by the fitting
function in \citet{Bryan1998}. 

For each halo in the simulation, the dimensionless spin parameter $\lambda$ is calculated 
by the formula \citep{Peebles1969}\footnote{An alternative definition is $\lambda= J/({\sqrt 2} MVR ) $ by \citet{Bullock2001}.} 
\begin{equation}\label{eqn:spin}
\lambda = \frac { J |E|^{1/2} }{G M_{\rm vir}^{5/2}},   
\end{equation}
where $J$ is the total angular momentum, $E$ is the total energy of the halo, and $G$ is
Newton’s gravitational constant. 

For a given halo with virial radius $R_{\rm vir}$, the concentration of the halo is defined
as $c_{\rm vir} = R_{\rm vir}/r_{\rm s}$,  where $r_{\rm s}$ is the scale radius in the density profiles 
fitted by the simple formula \citep{Navarro1997}
\begin{equation}\label{eqn:nfw}
\frac {\rho_{\rm NFW} (r)}{\rho_{\rm cirt}} = \frac {\delta_{\rm c}}{(r/r_{\rm s})(1 + r/r_{\rm s})^2} ,    
\end{equation}
where $\delta_{\rm c}$ is a characteristic density contrast, which can be expressed as 
$\delta_{\rm c} = \Delta_{\rm vir} c_{\rm vir}^3 /(3*(\ln(1+c_{\rm vir})-c_{\rm vir}(1+c_{\rm vir})))$
\citep{Maccio2007,Neto2007,Zhao2009}. Based on the haloes from the ELUCID simulation, the density 
profiles ${\rho (r_i)}$ is measured in $N$ spherical shells of uniform logarithmic bins from the 
radius where the bin contains at least 20 particles to $R_{\rm vir}$. The best fitting concentration
parameter is calculated by minimizing the rms deviation $\sigma$ between ${\rho (r_i)}$ and the density 
profile $\rho_{\rm NFW}(r_i)$ in Equation~\ref{eqn:nfw}
\begin{equation}\label{eqn:concentration}
\sigma^2 = \frac{1}{N} \sum_{i=1}^{N}   (\log {\rho (r_i)} - \log \rho_{\rm NFW}(r_i)  )^2.
\end{equation}

\subsection{Neighborhood subhalo abundance matching}\label{sec_SHAM}

Since the initial condition of the ELUCID simulation is constrained by the matter density
extracted from the galaxy distribution in SDSS DR7, the spatial distribution of the 
haloes in simulation is tightly correlated with the distribution of galaxies 
from SDSS DR7. Obviously, a set of gravitational bound subhaloes in simulation are 
more suitable to link galaxies in observation than FOF haloes. Therefore, we make use of 
a neighborhood subhalo abundance matching method \citep{Yang2018} to link galaxies in 
observation to subhaloes in the ELUCID simulation according to the likelihood
\begin{equation}\label{eqn:match} 
P = M_{\rm peak} \exp \left( - \frac {r_{p}^2} {2 r_{\rm off}^2} \right) 
\exp \left( -\frac {\pi^2} {2 v_{\rm off}^2}  \right), \end{equation}
where $r_{p}$ and $\pi$ are the separations between the galaxy and the subhalo in 
the perpendicular and parallel to the line-of-sight directions, respectively. $M_{\rm peak}$ 
is the subhalo peak mass over the merger histories. Compared with the current mass at $z=0$,  
the peak values are better traces of the potential well which shapes the galaxy statistical 
properties \citep{Reddick2013,Guo2016, Zhang2021b}. The free parameters $r_{\rm off}$ 
and  $v_{\rm off}$ are set to be $r_{\rm off} = 5 \mpc$ and $v_{\rm off} = 1000 ~{\rm km/s}$.
Note that $M_{\rm peak}$ is the dominant variable in Equation~\ref{eqn:match}, which will degrade 
to the traditional subhalo abundance matching method if $r_{\rm off} = \infty$ and $v_{\rm off} = \infty$.

In what follows, central galaxies are linked to central subhaloes in simulation and satellite
galaxies to satellite subhaloes. This matching criteria results in a total of $396,069$ 
galaxy-subhao pairs in the continuous Northern Galactic Cap (NGC) region of the range 
$99^{\circ} < \alpha < 283^{\circ}$ and $-7^{\circ} < \delta < 75^{\circ}$. Based on the 
galaxy-subhalo pairs in the matching catalog, the neighborhood subhalo abundance matching method
accurately reproduced the stellar to halo mass relation, the satellite fraction, the conditional 
stellar mass function, the biases of galaxies \citep{Yang2018}. In addition, the matching catalog
has been successfully applied to study the galaxy-halo alignments \citep{Zhang2021a}, and the physical connections 
between galaxy properties and halo formation time in different large-scale environments \citep{Zhang2021b}.

Unlike satellite galaxies which lose their mass due to tidal stripping after accretion into a larger system,  
central galaxies are more closely correlated with their host haloes in the ELUCID 
simulation \citep{Yang2018, Zhang2021a, Zhang2021b}. Therefore, in the following study, we mainly 
focus on the galaxy-halo size relation for a total of $276,463$ central galaxies.

\section{Galaxy-halo size relation}\label{sec_result}

Based on galaxies from SDSS DR7 and dark matter haloes from dark matter 
only cosmological ELUCID simulation, we investigate the galaxy-halo size 
relation for a total of $276,463$ central galaxies. In addition, we investigate
how the galaxy-halo size relation depends on the halo spin $\lambda$ and 
concentration $c_{\rm vir}$.

\subsection{Galaxy 2D half-light radius versus halo virial radius}

\begin{figure}
\includegraphics[width=0.5\textwidth]{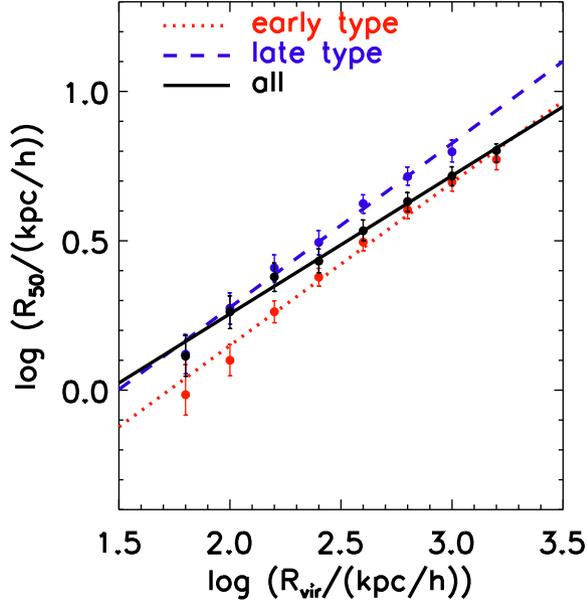}
\caption{Galaxy 2D half-light radius $R_{50}$ as a function of the host halo virial 
radius $R_{\rm vir}$. Galaxies are separated into early- and late-types according to 
the concentration index $c \ge 2.85$ and $c<2.85$. The red, blue and black symbols 
with error bars show the results for early-type, late-type and total galaxies, respectively. 
The black solid line shows the best-fitting relation $\log R_{50} = 0.46 
\log (R_{\rm vir}/ 210 \kpc) + 0.40$ for total galaxies. The red dotted lines 
shows the relation $\log R_{50} = 0.55 \log (R_{\rm vir}/210 \kpc) + 0.33$ for 
early-type galaxies, while the blue dashed line 
represents the best-fitting relation of the form $\log R_{50} = 0.55 
\log (R_{\rm vir}/210 \kpc) + 0.45$ for late-type galaxies.
}
\label{fig:Rvir_r50}
\end{figure}

At given stellar mass, galaxy size is strongly dependent on the specific star formation rate
\citep{VanderWel2014}. Star-forming galaxies are on average larger than quiescent galaxies at all redshifts \citep{Mowla2019}. Besides, the galaxy sizes
are also dependent on other properties of galaxies, such as concentration, morphology, and
bulge fraction \citep{Shen2003, Lange2015, Zhang2019, Irodotou2019}. The galaxy concentration index,
$c=R_{90}/R_{50}$, is tightly correlated with galaxy morphological types 
\citep{Deng2013, Deng2015, Calette2018}. Following \citet{Deng2013}, galaxies are 
separated into early- and late-types according to $c \ge 2.85$ and $c<2.85$, resulting in $78,238$ 
early-types and $198,225$ late-types.

Figure~\ref{fig:Rvir_r50} shows the projected 2D half-light radii $R_{50}$ of galaxies as a function
of their host halo 3D virial radii $R_{\rm vir}$.  The galaxy 2D half-light radii $R_{50}$ and halo 
virial radii $R_{\rm vir}$ are extracted from the NYU-VAGC catalog in observation and the halo catalog
in ELUCID simulation, respectively. The red, blue and black symbols with error bars 
show the results for early-type, late-type and total galaxies, respectively. Obviously, the galaxy-halo
size relation scales approximately linearly in logarithmic space over one order of magnitude in 
galaxy radius. For total galaxies in Figure~\ref{fig:Rvir_r50}, the black solid line 
shows the best-fitting relation $\log R_{50} = 0.46 \log (R_{\rm vir}/ 210 \kpc) + 0.40$ with an average
scatter $\langle \sigma \rangle = 0.14$ dex, where the average scatter 
$\langle \sigma \rangle$ is calculated by the mean absolute deviation from the best-fitting relation, and
$210 \kpc$ is the virial radius of the halo with mass $M_{\rm vir} = 10^{11.8} \msun$, which is the median 
value in our sample. For early-type galaxies, the best-fitting relation is 
$\log R_{50} = 0.55 \log (R_{\rm vir}/210 \kpc) + 0.33$ with  $\langle \sigma \rangle = 0.10$ dex 
indicated by the red dotted line, while for late-type galaxies, the galaxy-halo size relation is $\log R_{50} = 0.55 (\log R_{\rm vir}/210 \kpc) + 0.45$ with 
$\langle \sigma \rangle = 0.14$ dex shown by the blue dashed line. Generally, early- and late-type 
galaxies have almost the same power-law index of $0.55$ with $R_{50} \propto R_{\rm vir}^{0.55}$ and offset by $0.12$ dex.

\subsection{Galaxy 3D half-mass radius versus halo virial radius}

\begin{figure}
\includegraphics[width=0.5\textwidth]{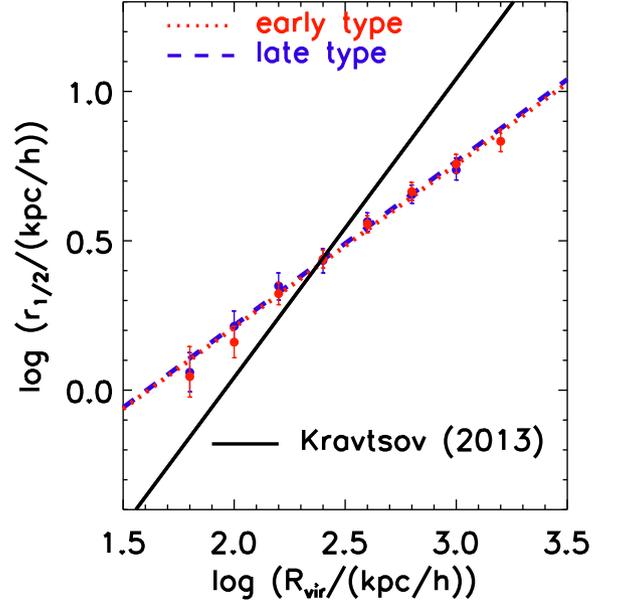}
\caption{Galaxy 3D half-mass radius $r_{1/2}$ as a function of the host halo virial 
radius $R_{\rm vir}$. The red and blue symbols with error bars show the results
for early- and late-type galaxies, respectively. The dotted and dashed lines shows 
the same best-fitting relation of the form $\log r_{1/2} = 0.55 \log (R_{\rm vir}/ 210 \kpc) + 0.39$ for
early- and late-type galaxies. The black solid line shows the galaxy-halo size relation obtained
by \citet{Kravtsov2013}.}
\label{fig:Rvir_r50_3D}
\end{figure}

\begin{figure}
\includegraphics[width=0.5\textwidth]{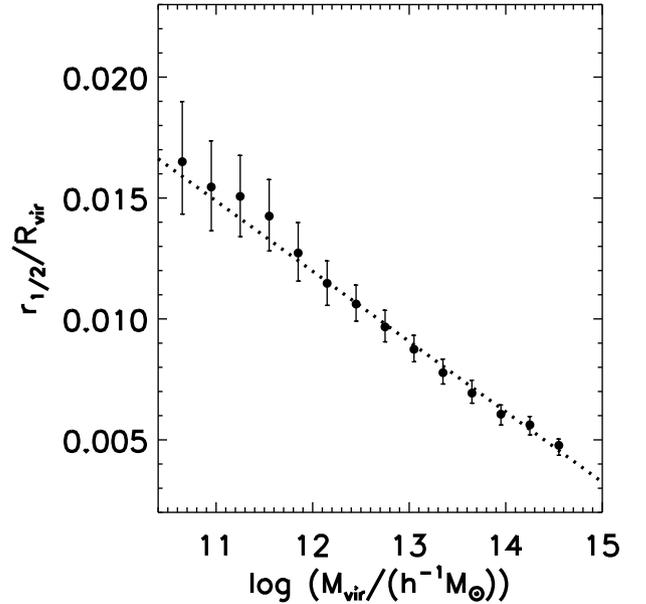}
\caption{Ratio of galaxy radius $r_{1/2}$ to halo virial radius $R_{\rm vir}$ as a function
of halo virial mass $M_{\rm vir}$. The solid dashed line shows the best-fitting relation
$r_{1/2}/R_{\rm vir} = 0.01\times(4.7-0.29  \log M_{\rm vir})$.
}
\label{fig:ratio_mvir}
\end{figure}

In this section, we investigate the size relation between 3D galaxy half-mass radius  $r_{1/2}$  and 
halo virial radius $R_{\rm vir}$. Inspired by \citet{Some2018}, the relation between 
the projected 2D half-light radius $R_{50}$ and the 3D half stellar mass radius $r_{1/2}$ is
expressed by
\begin{equation}\label{eqn:2d_3d}
r_{1/2} = f_{\rm i} f_{\rm m} R_{50},    
\end{equation}
where $f_{\rm i}$ corrects for the integration effect from the intrinsic properties in 3D space
to the projected distribution of the surface brightness \citep{Jaffe1983, Hernquist1990}, and $f_{\rm m}$
refers to the conversion from the half-light radius to the half-mass radius. 

The parameter $f_{\rm i}$ depends significantly on the shape and structure of galaxies 
\citep{Kravtsov2013, Some2018}. For spheroidal galaxies with s{\'e}rsic profiles, lots 
of studies have investigated the 3D density distributions for deprojecting the 2D density profiles 
in observation \citep{Lima1999, DeNicola2020, Vitral2020, Glenn2021}. Actually, there is no analytical
deprojection for the s{\'e}rsic profiles. In the numerical computations for spheroidal galaxies 
of de Vaucouleurs profiles with s{\'e}rsic index $n=4$, $f_{\rm i} = 1.35$ by integrating the 3D 
luminosity density along line-of-sight \citep{Young1976}. In the analytical model proposed 
by \citet{Hernquist1990}, $f_{\rm i} = 1.33$ by integrating 
the 3D spherical density distribution as the form $\rho(r) \propto r^{-2}(r+a)^{-2}$, 
which closely approximates the de Vaucouleurs profile in 2D density distribution. For spheroidal 
galaxies with a wide range of s{\'e}rsic index $n$, \citet{Lima1999} used a modified form of
the 3D density profile given by \citet{Mellier1987}, and derived the parameter 
$f_{\rm i} = 1.356 - 0.0293 \nu + 0.0023 \nu^2$, where $\nu = 1/n$, resulting in $f_{\rm i}= 1.35$ 
for de Vaucouleurs profile.

For disk galaxies, stars are assumed to be in a thin disk, and hence $f_{\rm i} = 1$ is adopted for late-type 
galaxies in the studies of \citet{Kravtsov2013} and \citet{Some2018}. In this study, early-type galaxies 
classified by galaxy concentration $c \ge 2.85$ are spheroid-dominated galaxies, while most of 
late-type galaxies with $c<2.85$ are fairly pure disk galaxies. In this paper, we adopt 
$f_{\rm i} = 1.32$ for early-type galaxies, and  $f_{\rm i} = 1$ for late-type galaxies.

The parameter $f_{\rm m}$ seems to be weakly dependent on the galaxy morphological types
\citep{Dutton2011, Szomoru2013, Lange2015, Chan2016, Suess2019a, Suess2019b, Ibarra2022}. 
Using hundreds of massive galaxies 
above $10^{10.7} {\rm M}_\odot$ from HST and SDSS observation, \citet{Szomoru2013} found that the half stellar mass 
radii are on average $25\%$ smaller than the $g$-band half-light radii and this difference of $25\%$ does not 
correlated with galaxy morphology or star forming activity. Using galaxies from GAMA survey 
in the redshift range $0.01<z<0.1$, \citet{Lange2015} found that both early- and late-type galaxies show 
a decrease in sizes of $13\%$ from $g$ to $K_s$ band at stellar mass $\sim 10^{10} {\rm M}_\odot$. The 
mass-weighted radius is assumed to be the same as the  $K_s$ band light effective radius, and then 
$f_{\rm m} = 0.87$ for the GAMA sample in \citet{Lange2015} . Using a sample of local early-type galaxies selected from the SPIDER survey, 
\citet{Chan2016} found that the mass-weighted sizes are on average $\sim 13\%$ smaller than the 
$r$-band sizes, corresponding to $f_{\rm m} = 0.87$ in the mass range 
$10.0 \leq  \log ({\rm M}_{*}/{\rm M}_\odot) \leq 11.6$  for the SPIDER sample. 

In this study, $f_{\rm m} = 0.87$ is adopted for early- or late-types, resulting in  
$f_{\rm i}f_{\rm m} = 1.32 * 0.87 = 1.15$ for early-types, and $f_{\rm i}f_{\rm m} = 1*0.87 = 0.87$ for late-types
in Equation~\ref{eqn:2d_3d}. 
Figure~\ref{fig:Rvir_r50_3D} show the galaxy 3D half-mass radius $r_{1/2}$ as a function of the halo virial 
radius $R_{\rm vir}$. The red and blue symbols with error bars show the results for early- and late-type galaxies,
respectively. The red dotted and blue dashed lines are the best-fitting galaxy-halo size relations for
early- and late-types. As shown in Figure~\ref{fig:Rvir_r50_3D}, there is no significant difference
of galaxy-halo size relations between early- and late-types in 3D space after correction for the 
integration effects.  Early- and late-type galaxies have the same best-fitting relation as the form 
$\log r_{1/2} = 0.55 \log (R_{\rm vir}/210 \kpc) + 0.39$ with the average scatter 
$\langle \sigma \rangle = 0.13$ dex.

It is noteworthy that the radius difference between early- and late-types is $0.12$ dex in the 2D case as shown 
in Figure~\ref{fig:Rvir_r50}. Therefore, the adopted values of $f_{\rm i} = 10^{0.12} = 1.32$ for early-types 
($f_{\rm i} = 1$ for late-types) can lead to the elimination of the radius difference in 3D space as shown in Figure~\ref{fig:Rvir_r50_3D}. 
We have repeated our entire calculation assuming $f_{\rm i} = 1.35$ for early-type galaxies, in which case
the 3D half-mass radius difference is $0.01$ dex between early- and late-type galaxies. Note that 
according to Equation~\ref{eqn:2d_3d}, the power-law index of $0.55$ is not changed from the 2D half-light radius $R_{50}$
to the 3D half-mass radius $r_{\rm 1/2}$ either for early-types or for late-types. 

However, there is an important caveat that the parameters $f_{\rm i}$ and $f_{\rm m}$ may be dependent on 
other galaxy properties beyond the morphological types in the conversion of 2D half-light radius to 3D
half-mass radius. Using $\sim 16500$ galaxies with stellar masses $10^{9.0} \msun \leq M_* \leq 10^{11.5} \msun$
in the CANDELS fields at $0 \leq z \leq 2.5$, \citet{Suess2019a,Suess2019b} investigated the galaxy
color gradients and quantified the ratios of half-mass to half-light radii at different redshifts. They found that
the ratios are dependent on the stellar mass or stellar mass surface density for early- or late-type galaxies
(see their Figure 5 of \citet{Suess2019a}). In addition, \citet{Suess2019b} claimed that the mass-to-light size 
ratios at fixed mass evolve rapidly with $1 \leq  z \leq 2.5$ , and remain roughly constant at $z \leq 1$. 
Using $537$ classical elliptical galaxies ($M_* \geq 10^{9} \msun$) 
from the MaNGA/SDSS-IV DR15 survey, \citet{Ibarra2022} also found the evolution of the mass-to-light
size ratio and  the dependence of the ratio on the galaxy stellar mass. 

In the analytical model for deprojecting s{\'e}rsic profiles for stellar systems of arbitrary triaxial shapes, 
\citet{Glenn2021} claimed that the projected semi-major radius is an unbiased proxy for the radius of a sphere 
that contains $50\%$ of the total luminosity or mass. Thus, we have also repeated our entire calculation using 
the semi-major radius from the catalog provided by \citet{Simard2011}. We find that the power-law index
$0.55$ is not changed and the overall scatter becomes slightly larger with 
$\langle \sigma \rangle = 0.17$ dex based on the semi-major radius instead of circular radius. However, in a previous
study, \citet{Kravtsov2013} found that the index of the relation between $r_{1/2}$ and $R_{\rm vir}$ is 
$\sim 1.0$, which is shown as the black solid line in Fig. \ref{fig:Rvir_r50_3D} for 
comparison\footnote{The best-fitting relation of \citet{Kravtsov2013} is 
$r_{1/2} = 0.015 R_{200} = 0.011 R_{\rm vir}$, where $R_{200}$ is the radius enclosing 
the over-density of $200$ times the critical density and $R_{\rm vir}$
is the halo virial radius defined by the fitting function used in \citet{Bryan1998}. 
The ratio is $R_{\rm vir}/R_{200} = 1.36 \pm 0.04$ according to the calculation
for dark matter haloes ($M_{\rm vir} \gtrsim 10^{10.5} \msun $) in the ELUCID simulation.}. This disagreement may be due to the difference of the detailed methodology in the abundance matching 
techniques \citep{Kravtsov2013, Yang2018}, where in estimating the halo mass for each galaxy, \cite{Kravtsov2013} adopted an abundance matching method without taking into account the scatter in the stellar-to-halo mass relation. 

Figure~\ref{fig:ratio_mvir} shows the galaxy-to-halo size ratio as a function of the halo virial mass.  
The solid dashed line shows the best-fitting relation $r_{1/2}/R_{\rm vir} = 0.01\times(4.7-0.29  \log M_{\rm vir})$.
Remarkably, the galaxy-halo size ratios depend significantly on the halo mass. Galaxies in more massive 
haloes have smaller ratios. In addition, we also calculate the galaxy-to-halo size ratio as a function of
the galaxy stellar mass $M_{*} \ga 10^9 \mstar$, and find that the relation can be approximated by the relation 
$r_{1/2}/R_{\rm vir} = 0.01 \times (4.8 - 0.35 \log M_{*})$. 

As shown in Figure~\ref{fig:Rvir_r50_3D}, the galaxy size $r_{1/2}$ is equal to that from the best-fitting relation
of \citet{Kravtsov2013} at fixed halo radius $R_{\rm vir} = 250 \kpc$, where the corresponding halo mass is
$M_{\rm vir} = 10^{12.3} \msun$ and $r_{1/2}/R_{\rm vir} = 0.011$. Note that the normalization factor is $0.015$
for $R_{200}$ in \citet{Kravtsov2013}, which corresponds to the factor $0.011$ for $R_{\rm vir}$ in this study.
As shown in Figure~\ref{fig:ratio_mvir}, the ratios of $r_{1/2}/R_{\rm vir}$ are smaller (larger) than $0.011$ for 
haloes more (less) than $10^{12.3} \msun$, therefore in Figure~\ref{fig:Rvir_r50_3D}, at fixed halo radii the 
galaxy sizes are smaller (larger) than those given by \citet{Kravtsov2013} for haloes more (less) than $10^{12.3} \msun$.
This trend leads to the power-law index of this study is significantly lower than that of \citet{Kravtsov2013}.
For haloes with mass larger than $\sim 10^{12} \msun$, the stellar-to-halo mass ratios decrease with increasing halo
mass \citep{Zhang2021b}, thus it's reasonable that the galaxy-halo size ratios are depdendent on the halo mass. 
In fact, the mass dependence of galaxy-halo size relation is also detected using the 2D galaxy size 
in \citet{Some2018}, although they claimed that the mass dependence is mitigated by the conversion from 2D half-light 
radius to 3D half-mass radius (see their Figure 3).

\subsection{halo spin dependence}
\begin{figure}
\includegraphics[width=0.24\textwidth, trim = 25 20 20 20]{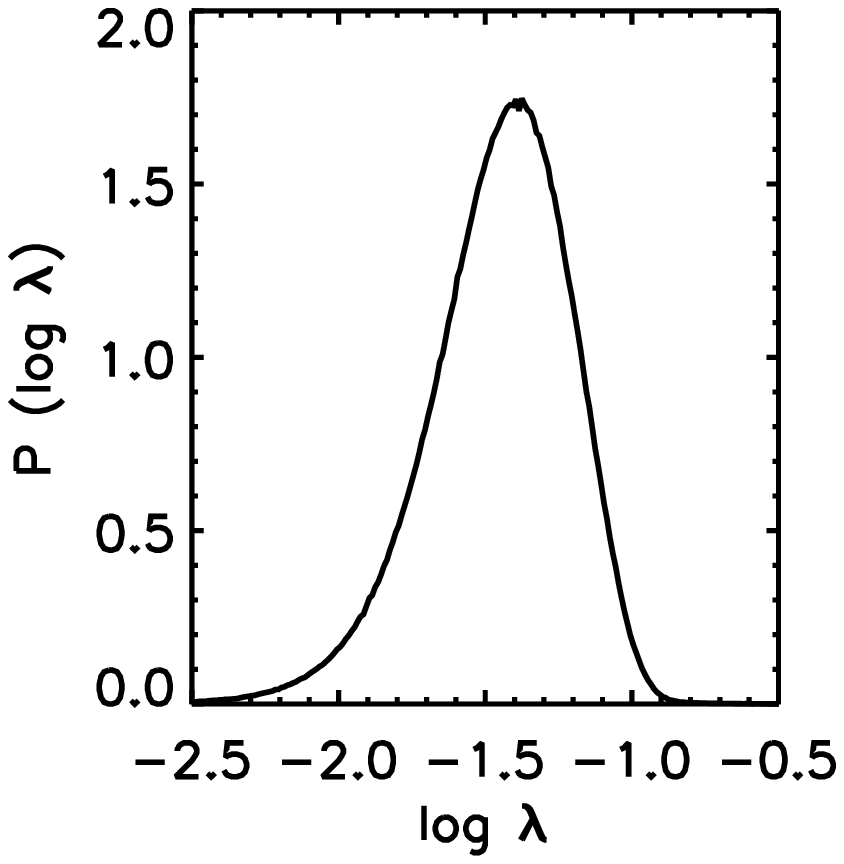}
\includegraphics[width=0.24\textwidth, trim = 25 20 20 20]{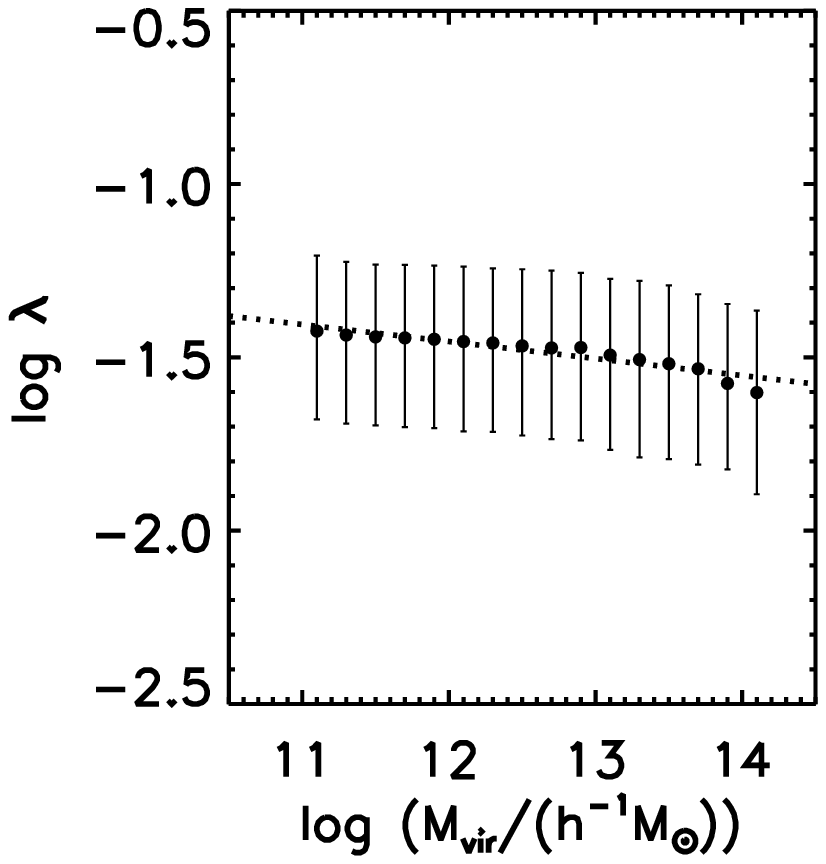}
\caption{Distribution of halo spin parameter $\lambda$  in the ELUCID simulation. 
\textbf{Left} panel: Probability distribution of the spin parameter $\lambda$ of dark matter 
haloes containing at least $300$ particles in the simulation. The median value 
of the spin parameter is $\langle \lambda \rangle =0.037$, and the standard deviation is  
$\sigma_{\log \lambda} = 0.24$. \textbf{Right} panel: Spin parameters as a function of 
the halo virial mass. The solid symbols are the median values, and the error bars show
the $16$ per cent and $84$ per cent quantiles.
}
\label{fig:spin}
\end{figure}

\begin{figure}
\includegraphics[width=0.5\textwidth]{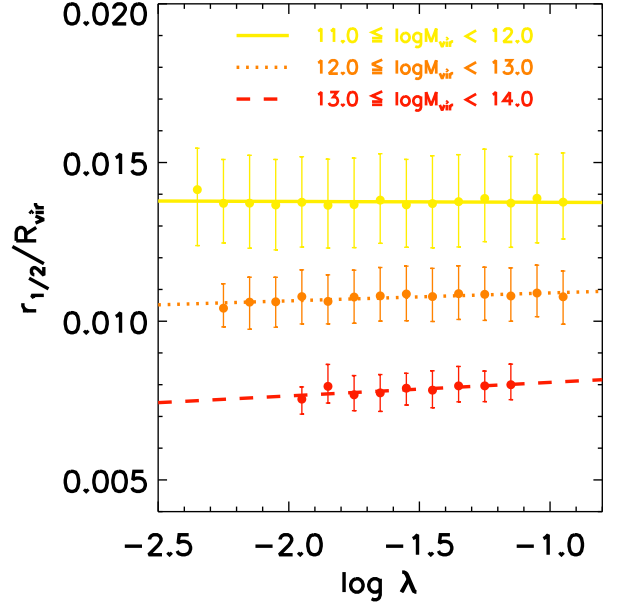}
\caption{Galaxy-to-halo size ratio as a function of the spin parameter with different mass
$\log (M_{\rm vir}/(\msun))$ in the ranges of $(11,12)$, $(12,13)$ and $(13,14)$. The solid 
symbols are the median values of $r_{1/2}/R_{\rm vir}$, and the error bars show the $40$ per cent and 
$60$ per cent quantiles. Different types of lines show the best fitting relations in different mass ranges. }
\label{fig:ratio_spin}
\end{figure}

In this section, we investigate how the galaxy-halo size relation depends on the halo spin.

The spin values are calculated using dark matter haloes from the ELUCID simulation according 
to Equation~\ref{eqn:spin}.
The left panel of Figure~\ref{fig:spin} shows the probability distribution of the spin parameter $\lambda$
of dark matter haloes with mass larger than $10^{11} \msun$, which contain at least $300$ particles in 
order to obtain a reliable measurement of halo properties, such as halo spin and concentration.
The median value of the spin parameter is $\langle \lambda \rangle =0.037$, and the standard deviation 
is  $\sigma_{\log \lambda} = 0.24$. Obviously, the median value and the scatter of the spin parameter are
in good agreement with previous works \citep{Bullock2001, Some2018, Jiang2019, YangH2021}.

The right panel of Figure~\ref{fig:spin} shows the spin parameters as a function of the halo virial mass.
The solid symbols are the median values, which can be fitted by the linear relation $\log \lambda = -0.049 
\log M_{\rm vir} - 0.865$. In simulation, the halo spin parameter is weakly dependent on halo 
mass, and larger haloes have slightly smaller spin parameters. Therefore, in the following analysis, galaxies
are separated into three subsamples with their corresponding halo mass $\log (M_{\rm vir}/(\msun))$ in the ranges of 
$(11,12)$, $(12,13)$ and $(13,14)$, in order to probe the spin dependence of the galaxy-to-halo size ratio.
As pointed out in \citet{Zhang2021b}, the matching between central galaxies and subhalos can be contaminated 
by nearby similar mass subhalos which may impact the link between the galaxy size and the spin parameters.
However, the matched galaxy-subhalo pairs are quite reliable for those subhaloes with mass $\ga 10^{12}\msun$.

Figure~\ref{fig:ratio_spin} shows galaxy-to-halo size ratio as a function of the spin parameter in different
mass ranges. The solid symbols with different colors are the median values of $r_{\rm 1/2}/R_{\rm vir}$. Different
types of lines shows the linear fitting functions in different mass ranges $(11,12)$, $(12,13)$ and $(13,14)$. 
Compared with the significant halo mass dependence, these results do not show any prominent correlation 
between $r_{\rm 1/2}/R_{\rm vir}$ and $\lambda$ at fixed mass ranges, including  the two 
massive bins where the matches are more reliable.  In general, these results are consistent with recent 
studies based on the hydrodynamical simulations \citep{Desmond2017,Jiang2019, YangH2021, Rohr2022, Stiskalek2022}.

Using two suites of zoom-in hydrodynamical simulation, \citet{Jiang2019} claimed that there is almost no  
correlation between galaxy-to-halo size ratio and the spin parameter, for $34$ galaxies with halo mass 
$\log (M_{\rm vir}/{\rm M}_\odot)$ in the mass range of $(11.3,12.3)$ from the VELA simulation, and for 
$\sim 100$ galaxies in the halo mass range $(9.5,12.3)$ from the NIHAO simulation, respectively. Using four
suits of cosmological hydrodynamical simulation, \citet{YangH2021} investigated the relation between 
$r_{1/2}/R_{\rm vir}$ and halo spin parameter $\lambda$. They found that for dwarfs with halo mass 
$\log (M_{200}/{\rm M}_\odot) \leq 11.2$, there is almost no correlation between 
$r_{\rm 1/2}/R_{\rm vir}$ and $\lambda$. For Milky Way sized galaxies in the mass range $(11.7,12.3)$, 
the correlation is very weak in the EAGLE simulation, although the correlation can be detected in the 
IllustrisTNG simulation. This indicates that different hydrodynamic solvers in different simulations 
could affect the relevant results. Using low-mass central galaxies ($M_* \sim 10^{7-9} M_\odot$) from 
the hydrodynamical FIREbox simulations, \citet{Rohr2022} found that the 
galaxy-halo size relation has almost no correlation with the halo spin, which agrees with the result in
this study.

In the canonical disk formation model of \citep{Fall1980,Mo1998}, it is assumed that the  
angular momenta of disk galaxies are fixed fractions of those of their surrounding haloes. Therefore,
in the angular momentum conservation model the galaxy spins are strongly correlated with those of their 
host haloes, resulting in the galaxy sizes are determined by the spins of their host haloes. However, 
in this study, we find that there is almost no correlation between galaxy size and the halo spin.
In addition, we have repeated our entire calculation for early- and late-type sub-samples, respectively.
There is also no correlation for early or late-type galaxies. Our finding disagrees with the canonical disk 
formation model in this regard.

\subsection{halo concentration dependence}

\begin{figure}
\includegraphics[width=0.5\textwidth]{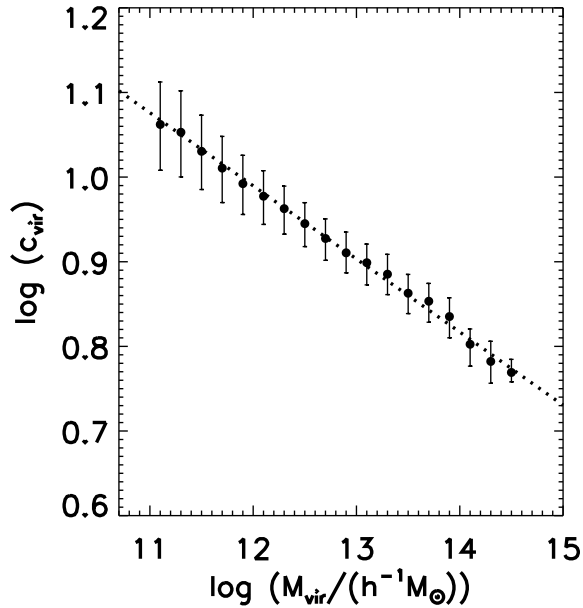}
\caption{Halo concentration $c_{\rm vir}$ as a function of halo virial mass in the ELUCID simulation. 
The solid symbols are the median values and the error bars show the $40$ per cent and 
$60$ per cent quantiles.
}
\label{fig:c_mvir}
\end{figure}

\begin{figure}
\includegraphics[width=0.5\textwidth]{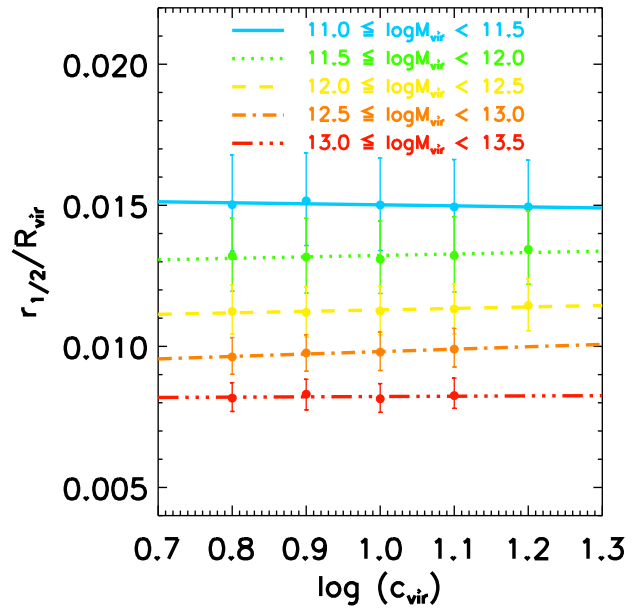}
\caption{Ratio of galaxy-to-halo size as a function of the halo concentration with different
mass in the ranges of $(11,11.5)$, $(11.5,12)$, $(12,12.5)$, $(12.5,13)$ and $(13,13.5)$.
The solid symbols are the median values and the error bars show the $40$ per cent and 
$60$ per cent quantiles.
}
\label{fig:ratio_c}
\end{figure}

In this section, we investigate how the galaxy-halo size relation depends on the halo concentration,
which has been widely used to describe the internal structure of haloes.

The halo concentration is defined as $c_{\rm vir} = R_{\rm vir}/r_{\rm s}$,  where $r_{\rm s}$ 
is the scale radius of the NFW profile given by Equation~\ref{eqn:nfw}.  The best-fitting values
are calculated by minimizing the rms deviation between the NFW density profiles and the halo 
density profiles from the ELUCID simulation according to Eqation~\ref{eqn:concentration}.
Figure~\ref{fig:c_mvir} shows the measured concentration $c_{\rm vir}$ as a function of 
the virial mass of the haloes matched to galaxies in observation. The solid symbols are the 
median values, which can be fitted by the relation $\log c_{\rm vir} = -0.09 
\log M_{\rm vir} + 2.03$, where the power-law index $-0.09$ is very close to the index $-0.1$ of the relaxed 
haloes in \citet{Neto2007}. 
Using $\sim 30$ million haloes ($M_{200} \geq 2 \times 10^{11} M_\odot $) of at 
least $2000$ particles from two very large dark matter only cosmological 
simulations, \citet{Child2018} investigated the concentration–mass relation in the redshift range
$0 \leq z \leq 4$. For haloes at $0 \leq z \leq 1$ fitted by the power-law relation 
(see their Table 2),  the power-law index $-0.089$ in \citet{Child2018}  agrees well with the 
index $-0.09$ in this study.

Obviously, the halo concentration is strongly dependent on the halo mass. Thus, in order to 
study the concentration dependence of galaxy-to-halo size ratio, galaxies are 
separated into five subsamples with their corresponding halo mass $\log (M_{\rm vir}/(\msun))$ 
in the ranges of $(11,11.5)$, $(11.5,12)$, $(12,12.5)$, $(12.5,13)$ and $(13,13.5)$. 
Figure~\ref{fig:ratio_c} shows the galaxy-to-halo size ratio as a function of the halo 
concentration in different mass ranges. The solid symbols with different colors are the median values,
and different types of lines show the best fitting linear relations in different mass ranges.
Based on these results, especially in the three high  massive ranges where the matches 
are more reliable, we do not see any prominent correlation between $r_{\rm 1/2}/R_{\rm vir}$ and
$c_{\rm vir}$ at fixed mass ranges. 

In this work, the independence of galaxy-halo size relation on halo concentration agrees 
with the results for low-mass central galaxies in the hydrodynamical simulation from the
FIRE project \citep{Rohr2022}, although \citet{Jiang2019} claimed that the galaxy-to-halo ratio is
dependent on the halo concentration with $r_{\rm 1/2}/R_{\rm vir} \propto c_{\rm vir}^{-0.7}$ based on the VELA 
and NIHAO simulation.

\section{Summary}\label{sec_summary}

Based on a sample of $276,463$ central galaxies ($M_* \gtrsim 10^{8} \mstar$) in SDSS observation and 
their corresponding haloes ($M_{\rm vir} \gtrsim 10^{10.5} \msun $) in the ELUCID simulation, we have 
investigated the galaxy-halo size relation in this study. Galaxy sizes are extracted from the NYU-VAGC 
catalog \citep{Blanton2005}, and the halo sizes are the virial radii ($R_{\rm vir}\sim 60-2000 \kpc$) of
dark matter haloes from the ELUCID simulation. Galaxies in observation are matched 
to (sub-)haloes in the constrained simulation using the neighborhood subhalo abundance matching
method \citep{Yang2018}. We then investigate the galaxy-halo size relation for central galaxies, 
which are more reliable than satellite galaxies in the matching pairs \citep{Zhang2021a, Zhang2021b}.
In addition, we investigate how the galaxy-halo size relation depends on the halo spin and 
concentration. The following are our main findings:

\begin{itemize}

\item [(i)]  Early-type galaxies have smaller 2D half-light radii $R_{50}$ than late-type galaxies at fixed halo virial 
radii $R_{\rm vir}$. They have the same \adb{power-law index} $0.55$ in the galaxy-halo size relation with $R_{50} \propto R_{\rm vir}^{0.55}$.

\item [(ii)]  There is no significant difference of galaxy 3D half-mass radius $r_{1/2}$  between early- and 
late-types. The vertical offset between early- and late-types is eliminated after correction from 
$R_{50}$ to $r_{1/2}$. The galaxy-halo size relation can be expressed as
$\log r_{1/2} = 0.55 \log (R_{\rm vir}/210 \kpc) + 0.39$. Note that the power-law 
index of $0.55$ is significantly lower than the values of $\sim 0.9 - 1.0$ in the previous 
studies \citep{Kravtsov2013, Some2018, Rohr2022}.

\item [(iii)] In this study, the galaxy-halo size ratios $r_{1/2}/R_{\rm vir}$ depend on the halo mass where galaxies in 
more massive haloes have smaller ratios, although \citet{Kravtsov2013} and  \citet{Some2018} claimed
that there is no mass dependence of the galaxy-halo size ratios. 

\item [(iv)] At fixed halo mass, there is no significant dependence of galaxy-to-halo size 
ratio on the halo spin or concentration, which agrees with recent works from hydrodynamical 
simulations \citep{Desmond2017,Jiang2019, YangH2021, Rohr2022}, but disagrees with the classical 
disk formation theory \citep{Fall1980,Mo1998}.

\end{itemize}

These results can provide guidelines for linking galaxy sizes to the sizes of their host haloes in the theoretical
and empirical models.

\section*{Acknowledgements}

We thank the anonymous referee for helpful comments that
significantly improve the presentation of this paper.
This work is supported by the national natural science foundation of China 
(Nos. 11833005,  11890692, 11621303), the CSST project 
No. CMS-CSST-2021-A02, 111 project No. B20019 and Shanghai Natural 
Science Foundation, grant No. 15ZR1446700. 

This work is also supported by the High Performance Computing Resource
in the Core Facility for Advanced Research Computing at Shanghai
Astronomical Observatory.

Funding for the Sloan Digital Sky Survey IV has been provided by the
Alfred P. Sloan Foundation, the U.S. Department of Energy Office of
Science, and the Participating Institutions. SDSS acknowledges support
and resources from the Center for High-Performance Computing at the
University of Utah. The SDSS web site is www.sdss.org.

SDSS is managed by the Astrophysical Research Consortium for the
Participating Institutions of the SDSS Collaboration including the
Brazilian Participation Group, the Carnegie Institution for Science,
Carnegie Mellon University, the Chilean Participation Group, the
French Participation Group, Harvard-Smithsonian Center for
Astrophysics, Instituto de Astrof{\'i}sica de Canarias, The Johns
Hopkins University, Kavli Institute for the Physics and Mathematics of
the Universe (IPMU)/University of Tokyo, Lawrence Berkeley National
Laboratory, Leibniz Institut f{\"u}r Astrophysik Potsdam (AIP),
Max-Planck-Institut f{\"u}r Astronomie (MPIA Heidelberg),
Max-Planck-Institut f{\"u}r Astrophysik (MPA Garching),
Max-Planck-Institut f{\"u}r Extraterrestrische Physik (MPE), National
Astronomical Observatories of China, New Mexico State University, New
York University, University of Notre Dame, Observat{\'o}rio Nacional/
MCTI, The Ohio State University, Pennsylvania State University,
Shanghai Astronomical Observatory, United Kingdom Participation Group,
Universidad Nacional Aut{\'o}noma de M{\'e}xico, University of
Arizona, University of Colorado Boulder, University of Oxford,
University of Portsmouth, University of Utah, University of Virginia,
University of Washington, University of Wisconsin, Vanderbilt
University, and Yale University.

\section*{Data availability}
The data underlying this article will be shared on reasonable request to the corresponding author.

\bibliographystyle{mnras}
\bibliography{bibliography}

\begin{thebibliography}{}
\makeatletter
\relax
\def\mn@urlcharsother{\let\do\@makeother \do\$\do\&\do\#\do\^\do\_\do\%\do\~}
\def\mn@doi{\begingroup\mn@urlcharsother \@ifnextchar [ {\mn@doi@}
  {\mn@doi@[]}}
\def\mn@doi@[#1]#2{\def\@tempa{#1}\ifx\@tempa\@empty \href
  {http://dx.doi.org/#2} {doi:#2}\else \href {http://dx.doi.org/#2} {#1}\fi
  \endgroup}
\def\mn@eprint#1#2{\mn@eprint@#1:#2::\@nil}
\def\mn@eprint@arXiv#1{\href {http://arxiv.org/abs/#1} {{\tt arXiv:#1}}}
\def\mn@eprint@dblp#1{\href {http://dblp.uni-trier.de/rec/bibtex/#1.xml}
  {dblp:#1}}
\def\mn@eprint@#1:#2:#3:#4\@nil{\def\@tempa {#1}\def\@tempb {#2}\def\@tempc
  {#3}\ifx \@tempc \@empty \let \@tempc \@tempb \let \@tempb \@tempa \fi \ifx
  \@tempb \@empty \def\@tempb {arXiv}\fi \@ifundefined
  {mn@eprint@\@tempb}{\@tempb:\@tempc}{\expandafter \expandafter \csname
  mn@eprint@\@tempb\endcsname \expandafter{\@tempc}}}

\bibitem[\protect\citeauthoryear{{Abazajian} et~al.,}{{Abazajian}
  et~al.}{2009}]{Abazajian2009}
{Abazajian} K.~N.,  et~al., 2009, \mn@doi [\apjs]
  {10.1088/0067-0049/182/2/543}, \href
  {https://ui.adsabs.harvard.edu/abs/2009ApJS..182..543A} {182, 543}

\bibitem[\protect\citeauthoryear{{Bell}, {McIntosh}, {Katz}  \&
  {Weinberg}}{{Bell} et~al.}{2003}]{Bell2003}
{Bell} E.~F.,  {McIntosh} D.~H.,  {Katz} N.,   {Weinberg} M.~D.,  2003, \mn@doi
  [\apjs] {10.1086/378847}, \href
  {https://ui.adsabs.harvard.edu/abs/2003ApJS..149..289B} {149, 289}

\bibitem[\protect\citeauthoryear{{Blanton} et~al.,}{{Blanton}
  et~al.}{2005}]{Blanton2005}
{Blanton} M.~R.,  et~al., 2005, \mn@doi [\aj] {10.1086/429803}, \href
  {https://ui.adsabs.harvard.edu/abs/2005AJ....129.2562B} {129, 2562}

\bibitem[\protect\citeauthoryear{{Bryan} \& {Norman}}{{Bryan} \&
  {Norman}}{1998}]{Bryan1998}
{Bryan} G.~L.,  {Norman} M.~L.,  1998, \mn@doi [\apj] {10.1086/305262}, \href
  {https://ui.adsabs.harvard.edu/abs/1998ApJ...495...80B} {495, 80}

\bibitem[\protect\citeauthoryear{{Bullock}, {Dekel}, {Kolatt}, {Kravtsov},
  {Klypin}, {Porciani}  \& {Primack}}{{Bullock} et~al.}{2001}]{Bullock2001}
{Bullock} J.~S.,  {Dekel} A.,  {Kolatt} T.~S.,  {Kravtsov} A.~V.,  {Klypin}
  A.~A.,  {Porciani} C.,   {Primack} J.~R.,  2001, \mn@doi [\apj]
  {10.1086/321477}, \href
  {https://ui.adsabs.harvard.edu/abs/2001ApJ...555..240B} {555, 240}

\bibitem[\protect\citeauthoryear{{Calette}, {Avila-Reese},
  {Rodr{\'\i}guez-Puebla}, {Hern{\'a}ndez-Toledo}  \& {Papastergis}}{{Calette}
  et~al.}{2018}]{Calette2018}
{Calette} A.~R.,  {Avila-Reese} V.,  {Rodr{\'\i}guez-Puebla} A.,
  {Hern{\'a}ndez-Toledo} H.,   {Papastergis} E.,  2018, \rmxaa, \href
  {https://ui.adsabs.harvard.edu/abs/2018RMxAA..54..443C} {54, 443}

\bibitem[\protect\citeauthoryear{{Chan} et~al.,}{{Chan}
  et~al.}{2016}]{Chan2016}
{Chan} J. C.~C.,  et~al., 2016, \mn@doi [\mnras] {10.1093/mnras/stw502}, \href
  {https://ui.adsabs.harvard.edu/abs/2016MNRAS.458.3181C} {458, 3181}

\bibitem[\protect\citeauthoryear{{Chang}, {van der Wel}, {da Cunha}  \&
  {Rix}}{{Chang} et~al.}{2015}]{Chang2015}
{Chang} Y.-Y.,  {van der Wel} A.,  {da Cunha} E.,   {Rix} H.-W.,  2015, \mn@doi
  [\apjs] {10.1088/0067-0049/219/1/8}, \href
  {https://ui.adsabs.harvard.edu/abs/2015ApJS..219....8C} {219, 8}

\bibitem[\protect\citeauthoryear{{Chen}, {Mo}, {Li}, {Wang}, {Yang}, {Zhou}  \&
  {Zhang}}{{Chen} et~al.}{2019}]{Chen2019}
{Chen} Y.,  {Mo} H.~J.,  {Li} C.,  {Wang} H.,  {Yang} X.,  {Zhou} S.,   {Zhang}
  Y.,  2019, \mn@doi [\apj] {10.3847/1538-4357/ab0208}, \href
  {https://ui.adsabs.harvard.edu/abs/2019ApJ...872..180C} {872, 180}

\bibitem[\protect\citeauthoryear{{Child}, {Habib}, {Heitmann}, {Frontiere},
  {Finkel}, {Pope}  \& {Morozov}}{{Child} et~al.}{2018}]{Child2018}
{Child} H.~L.,  {Habib} S.,  {Heitmann} K.,  {Frontiere} N.,  {Finkel} H.,
  {Pope} A.,   {Morozov} V.,  2018, \mn@doi [\apj] {10.3847/1538-4357/aabf95},
  \href {https://ui.adsabs.harvard.edu/abs/2018ApJ...859...55C} {859, 55}

\bibitem[\protect\citeauthoryear{{Cole}, {Lacey}, {Baugh}  \& {Frenk}}{{Cole}
  et~al.}{2000}]{Cole2000}
{Cole} S.,  {Lacey} C.~G.,  {Baugh} C.~M.,   {Frenk} C.~S.,  2000, \mn@doi
  [\mnras] {10.1046/j.1365-8711.2000.03879.x}, \href
  {https://ui.adsabs.harvard.edu/abs/2000MNRAS.319..168C} {319, 168}

\bibitem[\protect\citeauthoryear{{Croton} et~al.,}{{Croton}
  et~al.}{2006}]{Croton2006}
{Croton} D.~J.,  et~al., 2006, \mn@doi [\mnras]
  {10.1111/j.1365-2966.2005.09675.x}, \href
  {https://ui.adsabs.harvard.edu/abs/2006MNRAS.365...11C} {365, 11}

\bibitem[\protect\citeauthoryear{{Davis}, {Efstathiou}, {Frenk}  \&
  {White}}{{Davis} et~al.}{1985}]{Davis1985}
{Davis} M.,  {Efstathiou} G.,  {Frenk} C.~S.,   {White} S.~D.~M.,  1985,
  \mn@doi [\apj] {10.1086/163168}, \href
  {https://ui.adsabs.harvard.edu/abs/1985ApJ...292..371D} {292, 371}

\bibitem[\protect\citeauthoryear{{Deng}}{{Deng}}{2013}]{Deng2013}
{Deng} X.-F.,  2013, \mn@doi [Research in Astronomy and Astrophysics]
  {10.1088/1674-4527/13/6/004}, \href
  {https://ui.adsabs.harvard.edu/abs/2013RAA....13..651D} {13, 651}

\bibitem[\protect\citeauthoryear{{Deng} \& {Yu}}{{Deng} \&
  {Yu}}{2015}]{Deng2015}
{Deng} X.-F.,  {Yu} G.,  2015, \mn@doi [Astrophysics]
  {10.1007/s10511-015-9380-y}, \href
  {https://ui.adsabs.harvard.edu/abs/2015Ap.....58..250D} {58, 250}

\bibitem[\protect\citeauthoryear{{Desmond}, {Mao}, {Wechsler}, {Crain}  \&
  {Schaye}}{{Desmond} et~al.}{2017}]{Desmond2017}
{Desmond} H.,  {Mao} Y.-Y.,  {Wechsler} R.~H.,  {Crain} R.~A.,   {Schaye} J.,
  2017, \mn@doi [\mnras] {10.1093/mnrasl/slx093}, \href
  {https://ui.adsabs.harvard.edu/abs/2017MNRAS.471L..11D} {471, L11}

\bibitem[\protect\citeauthoryear{{Driver} et~al.,}{{Driver}
  et~al.}{2011}]{Driver2011}
{Driver} S.~P.,  et~al., 2011, \mn@doi [\mnras]
  {10.1111/j.1365-2966.2010.18188.x}, \href
  {https://ui.adsabs.harvard.edu/abs/2011MNRAS.413..971D} {413, 971}

\bibitem[\protect\citeauthoryear{{Dutton} \& {van den Bosch}}{{Dutton} \& {van
  den Bosch}}{2009}]{Dutton2009}
{Dutton} A.~A.,  {van den Bosch} F.~C.,  2009, \mn@doi [\mnras]
  {10.1111/j.1365-2966.2009.14742.x}, \href
  {https://ui.adsabs.harvard.edu/abs/2009MNRAS.396..141D} {396, 141}

\bibitem[\protect\citeauthoryear{{Dutton} et~al.,}{{Dutton}
  et~al.}{2011}]{Dutton2011}
{Dutton} A.~A.,  et~al., 2011, \mn@doi [\mnras]
  {10.1111/j.1365-2966.2011.19038.x}, \href
  {https://ui.adsabs.harvard.edu/abs/2011MNRAS.416..322D} {416, 322}

\bibitem[\protect\citeauthoryear{{Fall} \& {Efstathiou}}{{Fall} \&
  {Efstathiou}}{1980}]{Fall1980}
{Fall} S.~M.,  {Efstathiou} G.,  1980, \mn@doi [\mnras]
  {10.1093/mnras/193.2.189}, \href
  {https://ui.adsabs.harvard.edu/abs/1980MNRAS.193..189F} {193, 189}

\bibitem[\protect\citeauthoryear{{Garrison-Kimmel} et~al.,}{{Garrison-Kimmel}
  et~al.}{2018}]{Garrison2018}
{Garrison-Kimmel} S.,  et~al., 2018, \mn@doi [\mnras] {10.1093/mnras/sty2513},
  \href {https://ui.adsabs.harvard.edu/abs/2018MNRAS.481.4133G} {481, 4133}

\bibitem[\protect\citeauthoryear{{Grand} et~al.,}{{Grand}
  et~al.}{2017}]{Grand2017}
{Grand} R. J.~J.,  et~al., 2017, \mn@doi [\mnras] {10.1093/mnras/stx071}, \href
  {https://ui.adsabs.harvard.edu/abs/2017MNRAS.467..179G} {467, 179}

\bibitem[\protect\citeauthoryear{{Guo} et~al.,}{{Guo} et~al.}{2016}]{Guo2016}
{Guo} H.,  et~al., 2016, \mn@doi [\mnras] {10.1093/mnras/stw845}, \href
  {http://adsabs.harvard.edu/abs/2016MNRAS.459.3040G} {459, 3040}

\bibitem[\protect\citeauthoryear{{Hearin}, {Behroozi}, {Kravtsov}  \&
  {Moster}}{{Hearin} et~al.}{2019}]{Hearin2019}
{Hearin} A.,  {Behroozi} P.,  {Kravtsov} A.,   {Moster} B.,  2019, \mn@doi
  [\mnras] {10.1093/mnras/stz2251}, \href
  {https://ui.adsabs.harvard.edu/abs/2019MNRAS.489.1805H} {489, 1805}

\bibitem[\protect\citeauthoryear{{Hernquist}}{{Hernquist}}{1990}]{Hernquist1990}
{Hernquist} L.,  1990, \mn@doi [\apj] {10.1086/168845}, \href
  {https://ui.adsabs.harvard.edu/abs/1990ApJ...356..359H} {356, 359}

\bibitem[\protect\citeauthoryear{{Huang} et~al.,}{{Huang}
  et~al.}{2017}]{Huang2017}
{Huang} K.-H.,  et~al., 2017, \mn@doi [\apj] {10.3847/1538-4357/aa62a6}, \href
  {https://ui.adsabs.harvard.edu/abs/2017ApJ...838....6H} {838, 6}

\bibitem[\protect\citeauthoryear{{Ibarra-Medel}, {Avila-Reese}, {Lacerna},
  {Rodr{\'\i}guez-Puebla}, {V{\'a}zquez-Mata}, {Hern{\'a}ndez-Toledo}  \&
  {S{\'a}nchez}}{{Ibarra-Medel} et~al.}{2022}]{Ibarra2022}
{Ibarra-Medel} H.,  {Avila-Reese} V.,  {Lacerna} I.,  {Rodr{\'\i}guez-Puebla}
  A.,  {V{\'a}zquez-Mata} J.~A.,  {Hern{\'a}ndez-Toledo} H.~M.,   {S{\'a}nchez}
  S.~F.,  2022, \mn@doi [\mnras] {10.1093/mnras/stab3765}, \href
  {https://ui.adsabs.harvard.edu/abs/2022MNRAS.510.5676I} {510, 5676}

\bibitem[\protect\citeauthoryear{{Irodotou}, {Thomas}, {Henriques}, {Sargent}
  \& {Hislop}}{{Irodotou} et~al.}{2019}]{Irodotou2019}
{Irodotou} D.,  {Thomas} P.~A.,  {Henriques} B.~M.,  {Sargent} M.~T.,
  {Hislop} J.~M.,  2019, \mn@doi [\mnras] {10.1093/mnras/stz2365}, \href
  {https://ui.adsabs.harvard.edu/abs/2019MNRAS.489.3609I} {489, 3609}

\bibitem[\protect\citeauthoryear{{Jaffe}}{{Jaffe}}{1983}]{Jaffe1983}
{Jaffe} W.,  1983, \mn@doi [\mnras] {10.1093/mnras/202.4.995}, \href
  {https://ui.adsabs.harvard.edu/abs/1983MNRAS.202..995J} {202, 995}

\bibitem[\protect\citeauthoryear{{Jiang} et~al.,}{{Jiang}
  et~al.}{2019}]{Jiang2019}
{Jiang} F.,  et~al., 2019, \mn@doi [\mnras] {10.1093/mnras/stz1952}, \href
  {https://ui.adsabs.harvard.edu/abs/2019MNRAS.488.4801J} {488, 4801}

\bibitem[\protect\citeauthoryear{{Kravtsov}}{{Kravtsov}}{2013}]{Kravtsov2013}
{Kravtsov} A.~V.,  2013, \mn@doi [\apjl] {10.1088/2041-8205/764/2/L31}, \href
  {https://ui.adsabs.harvard.edu/abs/2013ApJ...764L..31K} {764, L31}

\bibitem[\protect\citeauthoryear{{Lange} et~al.,}{{Lange}
  et~al.}{2015}]{Lange2015}
{Lange} R.,  et~al., 2015, \mn@doi [\mnras] {10.1093/mnras/stu2467}, \href
  {https://ui.adsabs.harvard.edu/abs/2015MNRAS.447.2603L} {447, 2603}

\bibitem[\protect\citeauthoryear{{Lima Neto}, {Gerbal}  \& {M{\'a}rquez}}{{Lima
  Neto} et~al.}{1999}]{Lima1999}
{Lima Neto} G.~B.,  {Gerbal} D.,   {M{\'a}rquez} I.,  1999, \mn@doi [\mnras]
  {10.1046/j.1365-8711.1999.02849.x}, \href
  {https://ui.adsabs.harvard.edu/abs/1999MNRAS.309..481L} {309, 481}

\bibitem[\protect\citeauthoryear{{Macci{\`o}}, {Dutton}, {van den Bosch},
  {Moore}, {Potter}  \& {Stadel}}{{Macci{\`o}} et~al.}{2007}]{Maccio2007}
{Macci{\`o}} A.~V.,  {Dutton} A.~A.,  {van den Bosch} F.~C.,  {Moore} B.,
  {Potter} D.,   {Stadel} J.,  2007, \mn@doi [\mnras]
  {10.1111/j.1365-2966.2007.11720.x}, \href
  {https://ui.adsabs.harvard.edu/abs/2007MNRAS.378...55M} {378, 55}

\bibitem[\protect\citeauthoryear{{Meert}, {Vikram}  \& {Bernardi}}{{Meert}
  et~al.}{2015}]{Meert2015}
{Meert} A.,  {Vikram} V.,   {Bernardi} M.,  2015, \mn@doi [\mnras]
  {10.1093/mnras/stu2333}, \href
  {https://ui.adsabs.harvard.edu/abs/2015MNRAS.446.3943M} {446, 3943}

\bibitem[\protect\citeauthoryear{{Mellier} \& {Mathez}}{{Mellier} \&
  {Mathez}}{1987}]{Mellier1987}
{Mellier} Y.,  {Mathez} G.,  1987, \aap, \href
  {https://ui.adsabs.harvard.edu/abs/1987A&A...175....1M} {175, 1}

\bibitem[\protect\citeauthoryear{{Mo}, {Mao}  \& {White}}{{Mo}
  et~al.}{1998}]{Mo1998}
{Mo} H.~J.,  {Mao} S.,   {White} S. D.~M.,  1998, \mn@doi [\mnras]
  {10.1046/j.1365-8711.1998.01227.x}, \href
  {https://ui.adsabs.harvard.edu/abs/1998MNRAS.295..319M} {295, 319}

\bibitem[\protect\citeauthoryear{{Mowla} et~al.,}{{Mowla}
  et~al.}{2019}]{Mowla2019}
{Mowla} L.~A.,  et~al., 2019, \mn@doi [\apj] {10.3847/1538-4357/ab290a}, \href
  {https://ui.adsabs.harvard.edu/abs/2019ApJ...880...57M} {880, 57}

\bibitem[\protect\citeauthoryear{{Navarro}, {Frenk}  \& {White}}{{Navarro}
  et~al.}{1997}]{Navarro1997}
{Navarro} J.~F.,  {Frenk} C.~S.,   {White} S. D.~M.,  1997, \mn@doi [\apj]
  {10.1086/304888}, \href
  {https://ui.adsabs.harvard.edu/abs/1997ApJ...490..493N} {490, 493}

\bibitem[\protect\citeauthoryear{{Neto} et~al.,}{{Neto}
  et~al.}{2007}]{Neto2007}
{Neto} A.~F.,  et~al., 2007, \mn@doi [\mnras]
  {10.1111/j.1365-2966.2007.12381.x}, \href
  {https://ui.adsabs.harvard.edu/abs/2007MNRAS.381.1450N} {381, 1450}

\bibitem[\protect\citeauthoryear{{Peebles}}{{Peebles}}{1969}]{Peebles1969}
{Peebles} P.~J.~E.,  1969, \mn@doi [\apj] {10.1086/149876}, \href
  {https://ui.adsabs.harvard.edu/abs/1969ApJ...155..393P} {155, 393}

\bibitem[\protect\citeauthoryear{{Reddick}, {Wechsler}, {Tinker}  \&
  {Behroozi}}{{Reddick} et~al.}{2013}]{Reddick2013}
{Reddick} R.~M.,  {Wechsler} R.~H.,  {Tinker} J.~L.,   {Behroozi} P.~S.,  2013,
  \mn@doi [\apj] {10.1088/0004-637X/771/1/30}, \href
  {http://adsabs.harvard.edu/abs/2013ApJ...771...30R} {771, 30}

\bibitem[\protect\citeauthoryear{{Rodriguez-Gomez} et~al.,}{{Rodriguez-Gomez}
  et~al.}{2022}]{Rodriguez-Gomez2022}
{Rodriguez-Gomez} V.,  et~al., 2022, \mn@doi [\mnras] {10.1093/mnras/stac806},
  \href {https://ui.adsabs.harvard.edu/abs/2022MNRAS.512.5978R} {512, 5978}

\bibitem[\protect\citeauthoryear{{Rodriguez}, {Montero-Dorta}, {Angulo},
  {Artale}  \& {Merch{\'a}n}}{{Rodriguez} et~al.}{2021}]{Rodriguez2021}
{Rodriguez} F.,  {Montero-Dorta} A.~D.,  {Angulo} R.~E.,  {Artale} M.~C.,
  {Merch{\'a}n} M.,  2021, \mn@doi [\mnras] {10.1093/mnras/stab1571}, \href
  {https://ui.adsabs.harvard.edu/abs/2021MNRAS.505.3192R} {505, 3192}

\bibitem[\protect\citeauthoryear{{Rohr} et~al.,}{{Rohr}
  et~al.}{2022}]{Rohr2022}
{Rohr} E.,  et~al., 2022, \mn@doi [\mnras] {10.1093/mnras/stab3625}, \href
  {https://ui.adsabs.harvard.edu/abs/2022MNRAS.510.3967R} {510, 3967}

\bibitem[\protect\citeauthoryear{{Shen}, {Mo}, {White}, {Blanton}, {Kauffmann},
  {Voges}, {Brinkmann}  \& {Csabai}}{{Shen} et~al.}{2003}]{Shen2003}
{Shen} S.,  {Mo} H.~J.,  {White} S. D.~M.,  {Blanton} M.~R.,  {Kauffmann} G.,
  {Voges} W.,  {Brinkmann} J.,   {Csabai} I.,  2003, \mn@doi [\mnras]
  {10.1046/j.1365-8711.2003.06740.x}, \href
  {https://ui.adsabs.harvard.edu/abs/2003MNRAS.343..978S} {343, 978}

\bibitem[\protect\citeauthoryear{{Simard}, {Mendel}, {Patton}, {Ellison}  \&
  {McConnachie}}{{Simard} et~al.}{2011}]{Simard2011}
{Simard} L.,  {Mendel} J.~T.,  {Patton} D.~R.,  {Ellison} S.~L.,
  {McConnachie} A.~W.,  2011, \mn@doi [\apjs] {10.1088/0067-0049/196/1/11},
  \href {https://ui.adsabs.harvard.edu/abs/2011ApJS..196...11S} {196, 11}

\bibitem[\protect\citeauthoryear{{Somerville}, {Hopkins}, {Cox}, {Robertson}
  \& {Hernquist}}{{Somerville} et~al.}{2008}]{Some2008a}
{Somerville} R.~S.,  {Hopkins} P.~F.,  {Cox} T.~J.,  {Robertson} B.~E.,
  {Hernquist} L.,  2008, \mn@doi [\mnras] {10.1111/j.1365-2966.2008.13805.x},
  \href {https://ui.adsabs.harvard.edu/abs/2008MNRAS.391..481S} {391, 481}

\bibitem[\protect\citeauthoryear{{Somerville} et~al.,}{{Somerville}
  et~al.}{2018}]{Some2018}
{Somerville} R.~S.,  et~al., 2018, \mn@doi [\mnras] {10.1093/mnras/stx2040},
  \href {https://ui.adsabs.harvard.edu/abs/2018MNRAS.473.2714S} {473, 2714}

\bibitem[\protect\citeauthoryear{{Springel}, {White}, {Tormen}  \&
  {Kauffmann}}{{Springel} et~al.}{2001}]{Springel2001}
{Springel} V.,  {White} S. D.~M.,  {Tormen} G.,   {Kauffmann} G.,  2001,
  \mn@doi [\mnras] {10.1046/j.1365-8711.2001.04912.x}, \href
  {https://ui.adsabs.harvard.edu/abs/2001MNRAS.328..726S} {328, 726}

\bibitem[\protect\citeauthoryear{{Stiskalek}, {Bartlett}, {Desmond}  \&
  {Anbajagane}}{{Stiskalek} et~al.}{2022}]{Stiskalek2022}
{Stiskalek} R.,  {Bartlett} D.~J.,  {Desmond} H.,   {Anbajagane} D.,  2022,
  \mn@doi [\mnras] {10.1093/mnras/stac1609}, \href
  {https://ui.adsabs.harvard.edu/abs/2022MNRAS.514.4026S} {514, 4026}

\bibitem[\protect\citeauthoryear{{Suess}, {Kriek}, {Price}  \& {Barro}}{{Suess}
  et~al.}{2019a}]{Suess2019a}
{Suess} K.~A.,  {Kriek} M.,  {Price} S.~H.,   {Barro} G.,  2019a, \mn@doi
  [\apj] {10.3847/1538-4357/ab1bda}, \href
  {https://ui.adsabs.harvard.edu/abs/2019ApJ...877..103S} {877, 103}

\bibitem[\protect\citeauthoryear{{Suess}, {Kriek}, {Price}  \& {Barro}}{{Suess}
  et~al.}{2019b}]{Suess2019b}
{Suess} K.~A.,  {Kriek} M.,  {Price} S.~H.,   {Barro} G.,  2019b, \mn@doi
  [\apjl] {10.3847/2041-8213/ab4db3}, \href
  {https://ui.adsabs.harvard.edu/abs/2019ApJ...885L..22S} {885, L22}

\bibitem[\protect\citeauthoryear{{Szomoru}, {Franx}, {van Dokkum}, {Trenti},
  {Illingworth}, {Labb{\'e}}  \& {Oesch}}{{Szomoru} et~al.}{2013}]{Szomoru2013}
{Szomoru} D.,  {Franx} M.,  {van Dokkum} P.~G.,  {Trenti} M.,  {Illingworth}
  G.~D.,  {Labb{\'e}} I.,   {Oesch} P.,  2013, \mn@doi [\apj]
  {10.1088/0004-637X/763/2/73}, \href
  {https://ui.adsabs.harvard.edu/abs/2013ApJ...763...73S} {763, 73}

\bibitem[\protect\citeauthoryear{{Tweed}, {Yang}, {Wang}, {Cui}, {Zhang}, {Li},
  {Jing}  \& {Mo}}{{Tweed} et~al.}{2017}]{Tweed2017}
{Tweed} D.,  {Yang} X.,  {Wang} H.,  {Cui} W.,  {Zhang} Y.,  {Li} S.,  {Jing}
  Y.~P.,   {Mo} H.~J.,  2017, \mn@doi [\apj] {10.3847/1538-4357/aa6bf8}, \href
  {https://ui.adsabs.harvard.edu/abs/2017ApJ...841...55T} {841, 55}

\bibitem[\protect\citeauthoryear{{Vitral} \& {Mamon}}{{Vitral} \&
  {Mamon}}{2020}]{Vitral2020}
{Vitral} E.,  {Mamon} G.~A.,  2020, \mn@doi [\aap]
  {10.1051/0004-6361/201937202}, \href
  {https://ui.adsabs.harvard.edu/abs/2020A&A...635A..20V} {635, A20}

\bibitem[\protect\citeauthoryear{{Wang}, {Mo}, {Yang}  \& {van den
  Bosch}}{{Wang} et~al.}{2012}]{WangHuiyuan2012}
{Wang} H.,  {Mo} H.~J.,  {Yang} X.,   {van den Bosch} F.~C.,  2012, \mn@doi
  [\mnras] {10.1111/j.1365-2966.2011.20174.x}, \href
  {https://ui.adsabs.harvard.edu/abs/2012MNRAS.420.1809W} {420, 1809}

\bibitem[\protect\citeauthoryear{{Wang}, {Mo}, {Yang}, {Jing}  \& {Lin}}{{Wang}
  et~al.}{2014}]{WangHuiyuan2014}
{Wang} H.,  {Mo} H.~J.,  {Yang} X.,  {Jing} Y.~P.,   {Lin} W.~P.,  2014,
  \mn@doi [\apj] {10.1088/0004-637X/794/1/94}, \href
  {https://ui.adsabs.harvard.edu/abs/2014ApJ...794...94W} {794, 94}

\bibitem[\protect\citeauthoryear{{Wang} et~al.,}{{Wang}
  et~al.}{2016}]{WangHuiyuan2016}
{Wang} H.,  et~al., 2016, \mn@doi [\apj] {10.3847/0004-637X/831/2/164}, \href
  {https://ui.adsabs.harvard.edu/abs/2016ApJ...831..164W} {831, 164}

\bibitem[\protect\citeauthoryear{{Wang} et~al.,}{{Wang}
  et~al.}{2018}]{WangHY2018}
{Wang} H.,  et~al., 2018, \mn@doi [\apj] {10.3847/1538-4357/aa9e01}, \href
  {https://ui.adsabs.harvard.edu/abs/2018ApJ...852...31W} {852, 31}

\bibitem[\protect\citeauthoryear{{Wechsler} \& {Tinker}}{{Wechsler} \&
  {Tinker}}{2018}]{Wechsler2018}
{Wechsler} R.~H.,  {Tinker} J.~L.,  2018, \mn@doi [\araa]
  {10.1146/annurev-astro-081817-051756}, \href
  {https://ui.adsabs.harvard.edu/abs/2018ARA&A..56..435W} {56, 435}

\bibitem[\protect\citeauthoryear{{White} \& {Rees}}{{White} \&
  {Rees}}{1978}]{White1978}
{White} S.~D.~M.,  {Rees} M.~J.,  1978, \mn@doi [\mnras]
  {10.1093/mnras/183.3.341}, \href
  {https://ui.adsabs.harvard.edu/abs/1978MNRAS.183..341W} {183, 341}

\bibitem[\protect\citeauthoryear{{Yang} et~al.,}{{Yang}
  et~al.}{2018}]{Yang2018}
{Yang} X.,  et~al., 2018, \mn@doi [\apj] {10.3847/1538-4357/aac2ce}, \href
  {https://ui.adsabs.harvard.edu/abs/2018ApJ...860...30Y} {860, 30}

\bibitem[\protect\citeauthoryear{{Yang}, {Gao}, {Frenk}, {Grand}, {Guo}, {Liao}
   \& {Shao}}{{Yang} et~al.}{2021}]{YangH2021}
{Yang} H.,  {Gao} L.,  {Frenk} C.~S.,  {Grand} R. J.~J.,  {Guo} Q.,  {Liao} S.,
    {Shao} S.,  2021, arXiv e-prints, \href
  {https://ui.adsabs.harvard.edu/abs/2021arXiv211004434Y} {p. arXiv:2110.04434}

\bibitem[\protect\citeauthoryear{{Young}}{{Young}}{1976}]{Young1976}
{Young} P.~J.,  1976, \mn@doi [\aj] {10.1086/111959}, \href
  {https://ui.adsabs.harvard.edu/abs/1976AJ.....81..807Y} {81, 807}

\bibitem[\protect\citeauthoryear{{Zanisi} et~al.,}{{Zanisi}
  et~al.}{2020}]{Zanisi2020}
{Zanisi} L.,  et~al., 2020, \mn@doi [\mnras] {10.1093/mnras/stz3516}, \href
  {https://ui.adsabs.harvard.edu/abs/2020MNRAS.492.1671Z} {492, 1671}

\bibitem[\protect\citeauthoryear{{Zhang} \& {Yang}}{{Zhang} \&
  {Yang}}{2019}]{Zhang2019}
{Zhang} Y.-C.,  {Yang} X.-H.,  2019, \mn@doi [Research in Astronomy and
  Astrophysics] {10.1088/1674-4527/19/1/6}, \href
  {https://ui.adsabs.harvard.edu/abs/2019RAA....19....6Z} {19, 006}

\bibitem[\protect\citeauthoryear{{Zhang}, {Yang}  \& {Guo}}{{Zhang}
  et~al.}{2021a}]{Zhang2021a}
{Zhang} Y.,  {Yang} X.,   {Guo} H.,  2021a, \mn@doi [\mnras]
  {10.1093/mnras/staa2356}, \href
  {https://ui.adsabs.harvard.edu/abs/2021MNRAS.500.1895Z} {500, 1895}

\bibitem[\protect\citeauthoryear{{Zhang}, {Yang}  \& {Guo}}{{Zhang}
  et~al.}{2021b}]{Zhang2021b}
{Zhang} Y.,  {Yang} X.,   {Guo} H.,  2021b, \mn@doi [\mnras]
  {10.1093/mnras/stab2487}, \href
  {https://ui.adsabs.harvard.edu/abs/2021MNRAS.507.5320Z} {507, 5320}

\bibitem[\protect\citeauthoryear{{Zhao}, {Jing}, {Mo}  \& {B{\"o}rner}}{{Zhao}
  et~al.}{2009}]{Zhao2009}
{Zhao} D.~H.,  {Jing} Y.~P.,  {Mo} H.~J.,   {B{\"o}rner} G.,  2009, \mn@doi
  [\apj] {10.1088/0004-637X/707/1/354}, \href
  {https://ui.adsabs.harvard.edu/abs/2009ApJ...707..354Z} {707, 354}

\bibitem[\protect\citeauthoryear{{de Nicola}, {Saglia}, {Thomas}, {Dehnen}  \&
  {Bender}}{{de Nicola} et~al.}{2020}]{DeNicola2020}
{de Nicola} S.,  {Saglia} R.~P.,  {Thomas} J.,  {Dehnen} W.,   {Bender} R.,
  2020, \mn@doi [\mnras] {10.1093/mnras/staa1703}, \href
  {https://ui.adsabs.harvard.edu/abs/2020MNRAS.496.3076D} {496, 3076}

\bibitem[\protect\citeauthoryear{{van de Ven} \& {van der Wel}}{{van de Ven} \&
  {van der Wel}}{2021}]{Glenn2021}
{van de Ven} G.,  {van der Wel} A.,  2021, \mn@doi [\apj]
  {10.3847/1538-4357/abf047}, \href
  {https://ui.adsabs.harvard.edu/abs/2021ApJ...914...45V} {914, 45}

\bibitem[\protect\citeauthoryear{{van der Wel} et~al.,}{{van der Wel}
  et~al.}{2014}]{VanderWel2014}
{van der Wel} A.,  et~al., 2014, \mn@doi [\apj] {10.1088/0004-637X/788/1/28},
  \href {https://ui.adsabs.harvard.edu/abs/2014ApJ...788...28V} {788, 28}

\makeatother
\end{thebibliography}

% Don't change these lines
\bsp    % typesetting comment
\label{lastpage}
\end{document}